\documentstyle[12pt]{article}
\advance\voffset -8 mm
\advance\hoffset -1 cm
\def\alt{\mathrel{\mathpalette\vereq<}}
\def\agt{\mathrel{\mathpalette\vereq>}}
\def\vereq#1#2{\lower3pt\vbox{\baselineskip1.5pt \lineskip1.5pt
               \ialign{$\mpth#1\hfill##\hfil$\crcr#2\crcr\sim\crcr}}}
\def\mpth{\mathsurround=0pt}
\begin{document}
\pagestyle{plain}
\rightline{ JHU-TIPAC-920011 }
\rightline{ DFTT 12/92 }
\rightline{ March 1992 }
\vskip0.5cm
\centerline{\Large\bf Oscillations of Pseudo-Dirac Neutrinos }
\centerline{\Large\bf and the Solar Neutrino Problem }
\vskip0.5cm
\centerline{\Large C.~Giunti }
\centerline{\large\it Istituto Nazionale di Fisica Nucleare }
\centerline{\large\it Sezione di Torino }
\centerline{\large\it I--10125 Torino, Italy }
\vskip0.5cm
\centerline{\Large C.W.~Kim and U.W.~Lee }
\centerline{\large\it Department of Physics and Astronomy }
\centerline{\large\it The Johns Hopkins University }
\centerline{\large\it Baltimore, Maryland 21218, USA }
\vspace*{0.5in}
\centerline{\Large Abstract }
\bigskip
The oscillations of pseudo-Dirac neutrinos in matter are discussed
and applied to the solar neutrino problem. Several scenarios such as both
$\nu_e$ and $\nu_{\mu}$ being pseudo-Dirac and only $\nu_e$ or $\nu_{\mu}$
being pseudo-Dirac are examined. It is shown that the allowed region in the
mass-mixing angle parameter space obtained by comparing the solar neutrino
data with the calculations based on the standard solar model and
the MSW effect is not unique.
The results depend on the nature of neutrinos;
for example, if both $\nu_e$ and $\nu_{\mu}$ are pseudo-Dirac,
the allowed region
determined by the current solar neutrino data
does not overlap with that obtained
in the usual case of pure Dirac or Majorana neutrinos.

\newpage
The observation of the solar neutrinos, combined with the standard
solar model \cite{BAHCALL}
and calculations of the MSW effect \cite{MSW} can provide important
clues to understanding the basic properties of neutrinos.
Current data from Homestake \cite{CL},
Kamioka \cite{KII} and SAGE \cite{SAGE}
have already narrowed down considerably the allowed region in the
mass-mixing angle parameter space.

Recently there has been a proposal \cite{KLN91}
to explain the solar neutrino problem by
using the MSW effect with only one generation of pseudo-Dirac \cite{PSEU}
electron neutrinos with a large transition magnetic moment.
Majorana neutrinos emerge naturally in most
extensions of  the standard model,
but some models (see, for example, Ref.\cite{MODEL})
yield pseudo-Dirac neutrinos,
which were used, among others, to explain
the solar neutrino puzzle.

In this paper, we generalize the one generation picture to
the case of two generations and examine the consequences.
In particular, it is shown that in the pseudo-Dirac neutrino case
the current data yield entirely different (non-overlapping)
allowed regions in the mass-mixing angle parameter space from
those in the standard Dirac or Majorana neutrino cases.

The Dirac-Majorana mass matrix for the
neutrino states
$\nu^e_{\scriptscriptstyle L}$
and
$\nu^e_{\scriptscriptstyle R}$
is given by
\begin{equation}
\begin{array}{c|cc} \displaystyle
\quad \vphantom{\Big|} & \vphantom{\Big|} \quad
\overline{\nu^{e}_{\scriptscriptstyle L}}
\quad \vphantom{\Big|} & \vphantom{\Big|} \quad
\nu^{e}_{\scriptscriptstyle R}
\\ \hline \displaystyle
\nu^{e}_{\scriptscriptstyle L}
\quad \vphantom{\Big|} & \vphantom{\Big|} \quad
m^{e}_{\scriptscriptstyle L}
\quad \vphantom{\Big|} & \vphantom{\Big|} \quad
M^{e}_{\scriptscriptstyle D}
\\ \displaystyle
\overline{\nu^{e}_{\scriptscriptstyle R}}
\quad \vphantom{\Big|} & \vphantom{\Big|} \quad
M^{e}_{\scriptscriptstyle D}
\quad \vphantom{\Big|} & \vphantom{\Big|} \quad
m^{e}_{\scriptscriptstyle R}
\end{array}
\end{equation}
where $M^e_{\scriptscriptstyle D}$ and $m^e_{\scriptscriptstyle L,R}$
are Dirac and Majorana masses, respectively.
The assumption that
$ m^{e}_{\scriptscriptstyle L} + m^{e}_{\scriptscriptstyle R} = 0 $
leads to pseudo-Dirac neutrinos
generated by a mechanism similar to that
originally discussed by Wolfenstein \cite{PSEU}.
In the following we make the assumption that
$ M^e_{\scriptscriptstyle D} \gg | m^e_{\scriptscriptstyle L,R} | $
which also leads to pseudo-Dirac neutrinos \cite{Bilenky},
i.e.
two almost degenerate (in mass) left-handed neutrino
states
$\nu_1^e$ and $ \nu_2^e$
(with masses $m^e_1 \sim m^e_2 \sim M_{\scriptscriptstyle D}^e$)
which are expressed
as
\begin{equation}
\begin{array}{lll} \displaystyle \vphantom{\Bigg|}
\nu_1^e
& = & \displaystyle
  i\cos\theta_e \nu_{\scriptscriptstyle L}^e
- i\sin\theta_e \overline{\nu_{\scriptscriptstyle R}^e}
\\ \displaystyle \vphantom{\Bigg|}
\nu_2^e
& = & \displaystyle
  \sin\theta_e \nu_{\scriptscriptstyle L}^e
+ \cos\theta_e \overline{\nu_{\scriptscriptstyle R}^e}
\end{array}
\label{eq:002}
\end{equation}
where the factor $i$ guarantees the positivity of the mass eigenvalues.
The mixing angle $\theta_e$ is given by
\begin{equation}
\tan(2\theta_e) =
\frac{ 2 M^e_{\scriptscriptstyle D} }
     { ( m^e_{\scriptscriptstyle R} - m^e_{\scriptscriptstyle L} ) }
\nonumber
\label{eq:tan2theta}
\end{equation}
In our case
$\theta_e \sim 45^{\circ} $ because
$M_{\scriptscriptstyle D}^e \gg m_{\scriptscriptstyle L,R}^e$.
Therefore, our pseudo-Dirac neutrinos are special ones,
with almost $45^\circ$ mixing angle.
In general, however,
pseudo-Dirac neutrinos can have any mixing angle
and whenever pseudo-Dirac neutrinos are generated
with one generation one has to introduce sterile neutrinos.

In the following, we assume that
$ 10^{-11} \, {\rm eV}^2
  \alt \Delta m_e^2 \equiv (m_2^e)^2 - (m_1^e)^2 \alt
  10^{-7} \, {\rm eV}^2 $
where the upper limit comes from a cosmological constraint \cite{BD90}
on the oscillation into sterile neutrinos
and the lower limit is necessary in order to have
a vacuum oscillation length much shorter than the
sun-earth distance.
In this case, during the sun-earth propagation,
one half of the initial flux of $\nu_{\scriptscriptstyle L}^e$ will be
depleted due to the maximal ($45^{\circ}$ mixing)
oscillations between $\nu_{\scriptscriptstyle L}^e$ and
$\overline{\nu_{\scriptscriptstyle R}^e}$ when the time average is taken.
Therefore, the ratio of the $\nu^e_{\scriptscriptstyle L}$ flux
at the earth and the initial flux
is one half for the
SAGE (${\cal S}$), Homestake (${\cal H}$) and Kamioka (${\cal K}$)
experiments:
\begin{equation}
{\cal S} = {\cal H} = {\cal K} = \frac{1}{2}.
\label{eq:005}
\end{equation}

If neutrinos have magnetic moments large enough to
induce a spin-flip during their propagation in the magnetic
field of the sun \cite{VVO},
the ratios of the neutrino fluxes are
\begin{equation}
\nu^e_{\scriptscriptstyle L} : \overline{\nu^e_{\scriptscriptstyle L}} :
\nu^e_{\scriptscriptstyle R} : \overline{\nu^e_{\scriptscriptstyle R}}
=
\alpha : \frac{1-2 \alpha}{2} : \frac{1-2 \alpha}{2} : \alpha
\label{eq:006}
\end{equation}
with $0 \le \alpha \le 0.5$.
Since
$ \alpha $
is related to the spin-flip probability $P_{\rm sf}$ as
$ \alpha = (1-P_{\rm sf})/2 $,
any deviation from $\alpha=0.5$ is an indication that spin-flips
actually took place.
The detection rates are then
\begin{equation}
{\cal S} = {\cal H} = \alpha
\hskip0.25in , \hskip0.25in
{\cal K} = \alpha + 0.42 \, \frac{1-2\alpha}{2}
\label{eq:007}
\end{equation}
where we have used
$ \sigma(\overline{\nu^e_{\scriptscriptstyle L}} e^- ) \simeq
0.42 \sigma(\nu^e_{\scriptscriptstyle L} e^- ) $.
Assuming that the standard solar model \cite{BAHCALL}
gives the correct $\nu_{\scriptscriptstyle L}^e$ flux
produced in the core of the sun, the ratios of the observed fluxes
and the initial flux are
${\cal H}_{\rm exp} = 0.27 \pm 0.04 $ \cite{CL},
${\cal K}_{\rm exp} = 0.46 \pm 0.08 $ \cite{KII}
and
${\cal S}_{\rm exp} = 0.15 \pm 0.27 $ \cite{SAGE}.
The range of the parameter $\alpha$ for which Eq.(\ref{eq:007})
explains the observed ratios
is $ 0.29 \alt \alpha \alt 0.31$.

Now we generalize the above one generation picture to
the case of two generations and study its consequences.
We assume that,
before the mixing between the electron and muon sectors,
the muon neutrinos are also pseudo-Dirac particles,
i.e.
$M^{\mu}_{\scriptscriptstyle D} \gg | m^{\mu}_{\scriptscriptstyle L,R} | $
so that
$\theta_{\mu} \sim 45^{\circ}$.
The two almost degenerate
($ m^{\mu}_{\scriptscriptstyle 1} \sim
   m^{\mu}_{\scriptscriptstyle 2} \sim
   M^{\mu}_{\scriptscriptstyle D} $)
mass eigenstate muon-neutrino states
$\nu^{\mu}_{1}$ and
$\nu^{\mu}_{2}$ are given by
\begin{equation}
\begin{array}{lll} \displaystyle \vphantom{\Bigg|}
\nu_1^{\mu}
& = & \displaystyle
  i \cos\theta_{\mu} \nu^{\mu}_{\scriptscriptstyle L}
- i \sin\theta_{\mu} \overline{\nu^{\mu}_{\scriptscriptstyle R}}
\\ \displaystyle \vphantom{\Bigg|}
\nu_2^{\mu}
& = & \displaystyle
  \sin\theta_{\mu} \nu^{\mu}_{\scriptscriptstyle L}
+ \cos\theta_{\mu} \overline{\nu^{\mu}_{\scriptscriptstyle R}}
\end{array}
\label{eq:008}
\end{equation}
In the muon neutrino sector, we assume that
$ 10^{-11} {\rm eV}^2
  \alt \Delta m_{\scriptscriptstyle \mu}^2 \equiv
  (m^{\mu}_{2})^2 - (m^{\mu}_{1})^2 \alt
  10^{-2} {\rm eV}^2 $,
where the upper limit is due to the cosmological
constraint \cite{BD90}.
However, both the mixing angles among electron and muon neutrinos
and
$ \Delta m_{e\mu}^2 \sim
  (M_{\scriptscriptstyle D}^{\mu})^2 - (M_{\scriptscriptstyle D}^e)^2 $
are left unknown.
For simplicity, we assume that
$ \Delta m_{e\mu}^2 \gg \Delta m^2_e \, , \, \Delta m^2_{\mu} $
and that the mass eigenstates are given by
\begin{equation}
\left( \begin{array}{c} \nu_1 \\ \nu_2 \\ \nu_3 \\ \nu_4 \end{array} \right)
=
\left( \begin{array}{cccc}
 c_{\scriptscriptstyle \theta} & 0 & -s_{\scriptscriptstyle \theta} & 0 \\
 0 & c_{\scriptscriptstyle \theta} & 0 & -s_{\scriptscriptstyle \theta} \\
 s_{\scriptscriptstyle \theta} & 0 & c_{\scriptscriptstyle \theta} & 0 \\
 0 & s_{\scriptscriptstyle \theta} & 0 & c_{\scriptscriptstyle \theta}
       \end{array} \right)
{1\over\sqrt{2}}
\left( \begin{array}{cccc}
 i &-i & 0 & 0 \\
 1 & 1 & 0 & 0 \\
 0 & 0 & i &-i \\
 0 & 0 & 1 & 1
       \end{array} \right)
\left( \begin{array}{c}
\nu^e_{\scriptscriptstyle L} \\
\overline{\nu^e_{\scriptscriptstyle R}} \\
\nu^{\mu}_{\scriptscriptstyle L} \\
\overline{\nu^{\mu}_{\scriptscriptstyle R}}
\end{array} \right)
\label{eq:masseigen}
\end{equation}
where $ c_{\scriptscriptstyle\theta} \equiv \cos\theta_{e\mu} $
and   $ s_{\scriptscriptstyle\theta} \equiv \sin\theta_{e\mu} $.
The mixing angle $\theta_{e\mu}$ is practically equivalent to the usual mixing
angle between $\nu_e$ and $\nu_\mu$
in the Dirac or Majorana cases.
Also, in Eq.(\ref{eq:masseigen})
the pseudo-Dirac mixing angles $\theta_{e}$ and $\theta_{\mu}$ have been
approximated to $ 45^{\circ}$.
The mass matrix in the weak basis
$ M^{2}_{\scriptscriptstyle W} $
is given by
\begin{equation}
\hskip-1.0cm
M^{2}_{\scriptscriptstyle W} =
\left( \begin{array}{cccc}
\overline{m}^{2}_{12} c_{\scriptscriptstyle \theta}^2 + \overline{m}^{2}_{34}
s_{\scriptscriptstyle \theta}^2 + A_{e} &
\Delta m^2_{12} c_{\scriptscriptstyle \theta}^2
+ \Delta m^2_{34} s_{\scriptscriptstyle \theta}^2 &
\left( \overline{m}^{2}_{34} - \overline{m}^{2}_{12} \right)
c_{\scriptscriptstyle \theta}
s_{\scriptscriptstyle \theta} &
\left( \Delta m^2_{34} - \Delta m^2_{12} \right) c_{\scriptscriptstyle \theta}
s_{\scriptscriptstyle \theta} \vphantom{\Bigg|} \\
\Delta m^2_{12} c_{\scriptscriptstyle \theta}^2
+ \Delta m^2_{34} s_{\scriptscriptstyle \theta}^2 &
\overline{m}^{2}_{12} c_{\scriptscriptstyle \theta}^2 + \overline{m}^{2}_{34}
s_{\scriptscriptstyle \theta}^2 &
\left( \Delta m^2_{34} - \Delta m^2_{12} \right) c_{\scriptscriptstyle \theta}
s_{\scriptscriptstyle \theta} &
\left( \overline{m}^{2}_{34}
- \overline{m}^{2}_{12} \right) c_{\scriptscriptstyle \theta}
s_{\scriptscriptstyle \theta} \vphantom{\Bigg|} \\
\left( \overline{m}^{2}_{34}
- \overline{m}^{2}_{12} \right) c_{\scriptscriptstyle \theta}
s_{\scriptscriptstyle \theta} &
\left( \Delta m^2_{34} - \Delta m^2_{12} \right) c_{\scriptscriptstyle \theta}
s_{\scriptscriptstyle \theta} &
\overline{m}^{2}_{12} s_{\scriptscriptstyle \theta}^2 + \overline{m}^{2}_{34}
c_{\scriptscriptstyle \theta}^2 + A_{\mu} &
\Delta m^2_{12} s_{\scriptscriptstyle \theta}^2
+ \Delta m^2_{34} c_{\scriptscriptstyle \theta}^2 \vphantom{\Bigg|} \\
\left( \Delta m^2_{34} - \Delta m^2_{12} \right) c_{\scriptscriptstyle \theta}
s_{\scriptscriptstyle \theta} &
\left( \overline{m}^{2}_{34}
- \overline{m}^{2}_{12} \right) c_{\scriptscriptstyle \theta}
s_{\scriptscriptstyle \theta} &
\Delta m^2_{12} s_{\scriptscriptstyle \theta}^2
+ \Delta m^2_{34} c_{\scriptscriptstyle \theta}^2 &
\overline{m}^{2}_{12} s_{\scriptscriptstyle \theta}^2 + \overline{m}^{2}_{34}
c_{\scriptscriptstyle \theta}^2 \vphantom{\Bigg|}
\end{array} \right)
\end{equation}
with
\begin{equation}
\begin{array}{lll} \displaystyle \vphantom{\Bigg|}
\overline{m}^{2}_{12} \equiv { m_1^2 + m_2^2 \over 2 }
\ &,& \ \displaystyle
\overline{m}^{2}_{34} \equiv { m_3^2 + m_4^2 \over 2 }
\\ \displaystyle \vphantom{\Bigg|}
\Delta m^2_{12} \equiv { m_2^2 - m_1^2 \over 2 }
\ &,& \ \displaystyle
\Delta m^2_{34} \equiv { m_4^2 - m_3^2 \over 2 }
\\ \displaystyle \vphantom{\Bigg|}
A_{e} = A_{\scriptscriptstyle CC} + A_{\scriptscriptstyle NC}
\ &,& \ \displaystyle
A_{\mu} = A_{\scriptscriptstyle NC}
\\ \displaystyle \vphantom{\Bigg|}
A_{\scriptscriptstyle CC} =
2 \sqrt{2} \, G_{\scriptscriptstyle F} \, E \, N_{e}
\ &,& \ \displaystyle
A_{\scriptscriptstyle NC} =
- \sqrt{2} \, G_{\scriptscriptstyle F} \, E \, N_{n}
\end{array}
\end{equation}
where $m_{1} , m_{2} , m_{3}$ and $m_{4}$ are the mass eigenvalues.
The values of the effective mass squared in matter
are shown in Fig.1 for $\nu_1$, $\nu_2$, $\nu_3$ and $\nu_4$
and their antiparticles as functions of the matter density $\rho$.
In Fig.1 there are two MSW resonance regions
$R1$ and $R2$
and two possible spin-flip resonance regions
$R1_m$ and $R2_m$
in addition to the region $R$ discussed in Ref.\cite{KLN91}.
In the region $R$ the maximal vacuum oscillations
lead to the $1/2$ suppression of the $\nu^e_{\scriptscriptstyle L}$ flux,
as discussed above.

First we discuss the case in which neutrinos have large enough magnetic
moments to induce resonant spin-flips and the resonance
regions $R1_m$ and $R2_m$ are in the convection zone.
In this case the right-handed neutrinos
$\overline{\nu^e_{\scriptscriptstyle L}}$,
$\nu^e_{\scriptscriptstyle R}$,
$\overline{\nu^{\mu}_{\scriptscriptstyle L}}$
and
$\nu^{\mu}_{\scriptscriptstyle R}$
can be generated from the original
$\nu^e_{\scriptscriptstyle L}$
by the resonant spin-flip  processes directly or indirectly,
e.g. via
$ \nu^e_{\scriptscriptstyle L} \to
  \nu^{\mu}_{\scriptscriptstyle R} $,
$ \nu^e_{\scriptscriptstyle L} \to
  \overline{ \nu^{\mu}_{\scriptscriptstyle L}} $,
$ \nu^e_{\scriptscriptstyle L} \to
  \nu^e_{\scriptscriptstyle R} $,
$ \nu^e_{\scriptscriptstyle L} \to
  \nu^{\mu}_{\scriptscriptstyle R} \to
  \overline{ \nu^e_{\scriptscriptstyle L}} $.

Regardless of whether the resonance spin-flip processes are adiabatic
or not,
the flux ratios at the earth are expressed as
\begin{eqnarray}
\lefteqn{
\nu^e_{\scriptscriptstyle L} :
\overline{\nu^e_{\scriptscriptstyle L}} :
\nu^e_{\scriptscriptstyle R} :
\overline{\nu^e_{\scriptscriptstyle R}} :
\nu^{\mu}_{\scriptscriptstyle L} :
\overline{\nu^{\mu}_{\scriptscriptstyle L}} :
\nu^{\mu}_{\scriptscriptstyle R} :
\overline{\nu^{\mu}_{\scriptscriptstyle R}} =
}
\nonumber \\
&&
a\alpha : a \, \frac{1-2\alpha}{2} : a \, \frac{1-2\alpha}{2} : a \alpha :
b\beta : b \, \frac{1-2\beta}{2} : b \, \frac{1-2\beta}{2} : b \beta
\label{eq:015}
\end{eqnarray}
with $a + b = 1$ and $ 0 \le \alpha, \beta \le 0.5 $.
The detection rates are then
\begin{equation}
{\cal S} = {\cal H} = a \alpha \ , \hskip 0.5in
{\cal K} \simeq a \alpha + 0.21 a ( 1 - 2 \alpha)
+ \frac{1}{12}(1 - a)
\label{eq:016}
\end{equation}
where we have used
$ \sigma(\nu^{\mu}_{\scriptscriptstyle L} e^- )
   \simeq \frac{1}{6}
  \sigma(\nu^{e}_{\scriptscriptstyle L} e^-) $,
and
$ \sigma(\overline{\nu^{\mu}_{\scriptscriptstyle L}} e^-)
   \simeq \frac{1}{6}
  \sigma(\nu^{e}_{\scriptscriptstyle L} e^-) $
instead of
$ \sigma(\overline{\nu^{\mu}_{\scriptscriptstyle L}} e^-)
   \simeq \frac{1}{7}
  \sigma(\nu^{e}_{\scriptscriptstyle L} e^-) $
for simplicity.
This approximation makes the second equation in Eq.(\ref{eq:016}) free of
the parameter $\beta$.
The results in Eq.(\ref{eq:016}) are consistent with
${\cal S}_{\rm exp}$,
${\cal H}_{\rm exp}$ and
${\cal K}_{\rm exp}$
for the parameter ranges
$ a \agt 0.93$ and
$ 0.29 \alt \alpha \alt 0.31 $
within $1\sigma$.
The fact that $a$ must be very close to unity
implies that the Landau-Zener transitions
at the resonances $R1$ and $R2$
are extremely non-adiabatic in this model. Since the deviation from
$\alpha = 0.5$
indicates the presence of spin-flip,
the above range of $\alpha$
requires a large transition magnetic moment between
$\nu^e_{\scriptscriptstyle L}$ and
$\nu^e_{\scriptscriptstyle R}$
(e.g. $\sim 10^{-10} \mu_{B}$
for $B \sim 10 \, {\rm KG}$ in the convection zone).
Although there exist many models that can yield such a large magnetic moment,
they appear somewhat unnatural and thus we do not consider this scenario
further.

In the absence of magnetic moments, no right-handed neutrinos are produced
as $\nu_{\scriptscriptstyle L}^e$
($\simeq \nu_4$ in the core)
propagates outward from the core.
At the resonance R2, $\nu_4$ is split
into $\nu_4$ and $\nu_3$ with fractions
$(1-P_{\scriptscriptstyle R2})$ and $P_{\scriptscriptstyle R2}$,
respectively,
where
$P_{\scriptscriptstyle R2}$
is the Landau-Zener transition probability at the resonance R2.
At the resonance R1,
the fraction
$P_{\scriptscriptstyle R2}$ of
$\nu_3$ is further split
into $\nu_3$ and $\nu_2$ with fractions
$(1-P_{\scriptscriptstyle R1})P_{\scriptscriptstyle R2}$
and
$P_{\scriptscriptstyle R1}P_{\scriptscriptstyle R2}$,
respectively,
$P_{\scriptscriptstyle R1}$ being the Landau-Zener transition
probability at the resonance R1.

Let us assume that
the two resonance regions do not overlap so that
the two resonances can be treated separately.
Then one can estimate the respective
Landau-Zener transition probabilities as follows.

(1) {\sl First resonance} (R1).
This is a resonance between
$ \nu^{e}_{\scriptscriptstyle L} $ and
$ \nu^{\mu}_{\scriptscriptstyle L} $
and occurs when the first and third diagonal elements of
$ M^{2}_{\scriptscriptstyle W} $
are equal, i.e. for
$
A_{\scriptscriptstyle CC} =
\left( \overline{m}^{2}_{34} - \overline{m}^{2}_{12} \right)
c_{\scriptscriptstyle2\theta}.
$
In the neighborhood of the resonance the MSW evolution equation
is dominated by the
$ \nu^{e}_{\scriptscriptstyle L} $--$ \nu^{\mu}_{\scriptscriptstyle L} $
$ 2 \times 2 $ sector:
\begin{equation}
\left( \begin{array}{cc}
\overline{m}^{2}_{12} c_{\scriptscriptstyle\theta}^2 + \overline{m}^{2}_{34}
s_{\scriptscriptstyle\theta}^2 + A_{e} &
\left( \overline{m}^{2}_{34} - \overline{m}^{2}_{12} \right)
c_{\scriptscriptstyle\theta}
s_{\scriptscriptstyle\theta} \vphantom{\Bigg|}\\
\left( \overline{m}^{2}_{34} - \overline{m}^{2}_{12} \right)
c_{\scriptscriptstyle\theta}
s_{\scriptscriptstyle\theta} &
\overline{m}^{2}_{12} s_{\scriptscriptstyle\theta}^2 +
\overline{m}^{2}_{34}
c_{\scriptscriptstyle\theta}^2 + A_{\mu} \vphantom{\Bigg|}
\end{array} \right)
\ .
\end{equation}
The Landau-Zener transition probability at the resonance is given by
\begin{equation}
P_{\scriptscriptstyle R1} =
\exp\left[
- \frac{ \pi }{ 4 h_{\scriptscriptstyle R1} } \
\frac{ s_{\scriptscriptstyle2\theta}^2 }
     { c_{\scriptscriptstyle2\theta} } \
\frac{ \overline{m}^{2}_{34} - \overline{m}^{2}_{12} }{ E }
\right]
\ .
\label{EQ:CG23}
\end{equation}
with
$ h_{\scriptscriptstyle R1} \equiv
  \left. \frac{1}{\rho} \, \frac{\partial\rho}{\partial x}
  \right|_{\scriptscriptstyle R1} $.
The mass factor in the exponent,
$(\overline{m}^{2}_{34} - \overline{m}^{2}_{12})$,
is equivalent to the usual $\Delta m^2_{e\mu}$.

(2) {\sl Second resonance} (R2).
This is a resonance between
$ \nu^{e}_{\scriptscriptstyle L} $
and
$ \overline{\nu^{\mu}_{\scriptscriptstyle R}} $
and occurs when the first and fourth diagonal elements of
$ M^{2}_{\scriptscriptstyle W} $
are equal, i.e. for
$
A_{e} = \left( \overline{m}^{2}_{34} - \overline{m}^{2}_{12} \right)
c_{\scriptscriptstyle2\theta}
$.
In the neighborhood of the resonance the MSW evolution equation
is dominated by the
$ \nu^{e}_{\scriptscriptstyle L}
$--$
  \overline{\nu^{\mu}_{\scriptscriptstyle R}} $
$ 2 \times 2 $ sector:
\begin{equation}
M^{2}_{\scriptscriptstyle W} =
\left( \begin{array}{cc}
\overline{m}^{2}_{12} c_{\scriptscriptstyle\theta}^2 + \overline{m}^{2}_{34}
s_{\scriptscriptstyle\theta}^2 + A_{e} &
\left( \Delta m^2_{34} - \Delta m^2_{12} \right) c_{\scriptscriptstyle\theta}
s_{\scriptscriptstyle\theta} \vphantom{\Bigg|}\\
\left( \Delta m^2_{34} - \Delta m^2_{12} \right) c_{\scriptscriptstyle\theta}
s_{\scriptscriptstyle\theta} &
\overline{m}^{2}_{12} s_{\scriptscriptstyle\theta}^2 + \overline{m}^{2}_{34}
c_{\scriptscriptstyle\theta}^2 \vphantom{\Bigg|}
\end{array} \right)
\ .
\end{equation}
The Landau-Zener transition probability at the resonance is given by
\begin{equation}
P_{\scriptscriptstyle R2} =
\exp\left[
- \frac{ \pi }{ 4 h_{\scriptscriptstyle R2} } \
\frac{ s_{\scriptscriptstyle2\theta}^2 }
     { c_{\scriptscriptstyle2\theta} } \
\frac{ \overline{m}^{2}_{34} - \overline{m}^{2}_{12} }{ E } \
\left(
\frac{ \Delta m^2_{34} - \Delta m^2_{12} }
     { \overline{m}^{2}_{34} - \overline{m}^{2}_{12} }
\right)^2
\right]
\ .
\label{EQ:CG28}
\end{equation}
with
$
h_{\scriptscriptstyle R2} \equiv
\left. \frac{1}{\rho} \, \frac{\partial\rho}{\partial x}
\right|_{\scriptscriptstyle R2}
$.
The ratio of the mass factors in the exponents of
Eqs.(\ref{EQ:CG23}) and (\ref{EQ:CG28}) is
$
[ (\Delta m_{34}^2 - \Delta m_{12}^2 )
   /
  (\overline{m}_{34}^2 - \overline{m}_{12}^2 )  ]^2
\sim
(\Delta m_{34}^2 / \overline{m}_{34}^2  )^2
\sim
( M_{\scriptscriptstyle D}^{\mu} m^{\mu}_{\scriptscriptstyle L,R} /
  {M_{\scriptscriptstyle D}^{\mu}}^2  )^2
\sim
( m^{\mu}_{\scriptscriptstyle L,R} / M_{\scriptscriptstyle D}^{\mu}  )^2$.
The numerical value of this ratio is supposed to be small for the
pseudo-Dirac neutrinos under consideration.
Therefore, we take the non-adiabatic
approximation for the resonant transition at $R2$.
In order to see the region of validity for this approximation,
let us consider the exponent of Eq.(\ref{EQ:CG28}) which is written as
\begin{eqnarray}
     Q_{R2}
     &\simeq &
- \frac{\pi}{4}
  \frac{ \sin^2(2\theta_{e\mu}) }
       { \cos(2\theta_{e\mu}) }
  \frac{R_{\odot}}{10.45} \frac{\Delta m^2_{e\mu}}{E}
  \left(
  \frac{m^{\mu}_{\scriptscriptstyle L,R}}
       {M_{\scriptscriptstyle D}^{\mu}}
  \right)^2
  \nonumber  \\
      &\simeq &
- 2.6 \times 10^{3} \
  \frac{ \sin^2(2\theta_{e\mu}) }
       { \cos(2\theta_{e\mu}) } \
  \frac{ \Delta m^2_{e\mu} }
       { {\rm eV}^2 }
\end{eqnarray}
where we have used $E= 10$ MeV and for definiteness
$ m^{\mu}_{\scriptscriptstyle L,R} /
  M_{\scriptscriptstyle D}^{\mu}
  \simeq 0.01 $
which corresponds to the mixing angle
$\theta_{\mu} = 44.86^{\circ}$.
The non-adiabatic region which satisfies $ | Q_{R2} | < 1 $ is
below the solid line in the upper right-hand corner in Fig.2.
It will be shown that the solution
of the solar neutrino problem in the pseudo-Dirac neutrino model
indeed lies in this region.

Therefore, one has the following ratios of fluxes at the earth
\begin{equation}
\nu^e_{\scriptscriptstyle L} : \nu^{\mu}_{\scriptscriptstyle L} =
\frac{1}{2}
\Big[
P_{\scriptscriptstyle R1} \cos^2\theta_{e\mu} +
( 1 - P_{\scriptscriptstyle R1} )\sin^2\theta_{e\mu}
\Big] :
\frac{1}{2}
\Big[
P_{\scriptscriptstyle R1} \sin^2\theta_{e\mu}  +
( 1 - P_{\scriptscriptstyle R1} ) \cos^2\theta_{e\mu}
\Big]
\label{eq:012a}
\end{equation}
leading to
\begin{equation}
{\cal S} = {\cal H} = \frac{1}{2}\left[\frac{1}{2} - \frac{1}{2}
(1-2P_{\scriptscriptstyle R1} )\cos(2\theta_{e\mu}) \right]
\hskip0.25in , \hskip0.25in
{\cal K} = \frac{1}{12} + \frac{5}{12}
      \left[\frac{1}{2} - \frac{1}{2}
(1-2P_{\scriptscriptstyle R1} )\cos(2\theta_{e\mu}) \right]
\label{eq:013a}
\end{equation}
where we have used the fact that, because of
the maximal mixings between
$\nu^e_{\scriptscriptstyle L} \leftrightarrow
\overline{\nu^e_{\scriptscriptstyle R}}$ and
$\nu^{\mu}_{\scriptscriptstyle L} \leftrightarrow
\overline{\nu^{\mu}_{\scriptscriptstyle R}}$,
only half of the neutrinos can be detected.

Treating $P_{\scriptscriptstyle R1}$
as a constant parameter (neglecting the energy dependence
of the transition probability),
${\cal S}_{\rm exp}$,
${\cal H}_{\rm exp}$ and
${\cal K}_{\rm exp}$
are not reproduced by
Eq.(\ref{eq:013a})
within $1\sigma$ for any values of $P_{\scriptscriptstyle R1}$,
but are reproduced within $2\sigma$ for
\begin{equation}
0.52 \alt
\left[ \frac{1}{2}
- \frac{1}{2}( 1-2P_{\scriptscriptstyle R1} )\cos(2\theta_{e\mu})
\right]
\alt 0.70
\label{eq:013aa}
\end{equation}

The two-generation pseudo-Dirac neutrinos
discussed here are a special case of two-generation sterile neutrinos
in which $\theta_e$ and $\theta_\mu$ are $45^\circ$.
Since the only mixing angle which is relevant in the analysis
is $\theta_{e\mu}$, the above result should be compared with
the usual MSW effect with two
generations of neutrinos with the same mixing angle.
In this case one has
\begin{equation}
{\cal S} = {\cal H} = \frac{1}{2}
- \frac{1}{2}(1-2P_{\scriptscriptstyle\rm LZ})\cos(2\theta_{e\mu})
\hskip 0.25in , \hskip 0.25in
{\cal K} = \frac{1}{6} + \frac{5}{6}
\left[ \frac{1}{2}
- \frac{1}{2}(1-2P_{\scriptscriptstyle\rm LZ})\cos(2\theta_{e\mu})
\right]
\label{eq:014}
\end{equation}
with the Landau-Zener transition probability $P_{\scriptscriptstyle\rm LZ}$.
Equation (\ref{eq:014}) can reproduce
${\cal S}_{\rm exp}$,
${\cal H}_{\rm exp}$ and
${\cal K}_{\rm exp}$ within $1\sigma$ if
$ \displaystyle
0.26 \alt
\left[ \frac{1}{2}
- \frac{1}{2}(1-2P_{\scriptscriptstyle\rm LZ})\cos(2\theta_{e\mu})
\right]
\alt 0.31
$,
and within $2\sigma$ if
\begin{equation}
0.19 \alt
\left[ \frac{1}{2}
- \frac{1}{2}(1-2P_{\scriptscriptstyle\rm LZ})\cos(2\theta_{e\mu})
\right]
\alt 0.35
\ .
\label{eq:ordmsw}
\end{equation}

There is an important difference between
Eqs.(\ref{eq:013a}) and (\ref{eq:014}):
In Eq.(\ref{eq:013a}),
there is
an additional factor $1/2$ due to the pseudo-Dirac nature of neutrinos,
i.e.
$\nu^e_{\scriptscriptstyle L}$ and
$\nu^{\mu}_{\scriptscriptstyle L}$
oscillate into
$\overline{\nu^e_{\scriptscriptstyle R}}$ and
$\overline{\nu^{\mu}_{\scriptscriptstyle R}}$,
respectively,
with $45^{\circ}$ mixing,
depleting the  active (to detection) neutrinos by one half.

Since $P_{\scriptscriptstyle R1}$
depends on
the energy $E$
and the SAGE, Homestake and Kamioka experiments
have different energy thresholds,
$P_{\scriptscriptstyle R1}$
can be different for
${\cal S}$, ${\cal H}$ and ${\cal K}$.
Using the expression for $P_{\scriptscriptstyle R1}$,
we have plotted  the
$\Delta m_{e\mu}^2$ -- $\sin^2(2\theta_{e\mu})/\cos(2\theta_{e\mu})$
diagram in Fig.2
for the pseudo-Dirac MSW and the usual MSW effects based on
${\cal S}_{\rm exp}$, ${\cal H}_{\rm exp}$ and ${\cal K}_{\rm exp}$.
We have taken $\langle E \rangle = 2.0$ MeV,
$7.5$ MeV and $10$ MeV for
${\cal S}$,
${\cal H}$ and ${\cal K}$,  respectively.
The region which satisfies ${\cal S}_{\rm exp}$, ${\cal H}_{\rm exp}$ and
${\cal K}_{\rm exp}$ within $2\sigma$ in the standard two generation MSW
effect (with Eq.(\ref{eq:014})) is shown as
the area inside the dotted lines in Fig.2.
The region which satisfies ${\cal S}_{\rm exp}$, ${\cal H}_{\rm exp}$ and
${\cal K}_{\rm exp}$ within $2\sigma$ in the pseudo-Dirac case
(with Eq.(\ref{eq:013a})) is shown as
the shaded area inside the solid lines in Fig.2.
It is important to emphasize here that the two allowed
regions in the
$\Delta m_{e\mu}^2$ -- $\sin^2(2\theta_{e\mu})/\cos(2\theta_{e\mu} )$
plot based on the same data,
${\cal S}_{\rm exp}$, ${\cal H}_{\rm exp}$ and ${\cal K}_{\rm exp}$, do not
overlap,
even within $2\sigma$ errors,
i.e. the two generation pseudo-Dirac neutrinos produce
different allowed regions from those based on the usual two generation model.
As variations of the above pseudo-Dirac neutrino scenario, we consider
the following two cases:
(I) The electron neutrino is a pseudo-Dirac neutrino
but the muon neutrino is an ordinary Majorana or Dirac neutrino;
(II) The electron neutrino is a Majorana or Dirac neutrino but the muon
neutrino is a pseudo-Dirac neutrino.
In case (I) there is one MSW resonance region and
the neutrino fluxes at the earth
depend on the corresponding Landau-Zener transition
probability $P_{\scriptscriptstyle R1}$.
By assuming a mixing
between $\nu_{\mu}$ and $\nu_2^e$
with mixing angle $\theta_{e\mu} $, we have
\begin{equation}
\nu^e_{\scriptscriptstyle L} : \nu^{\mu}_{\scriptscriptstyle L} =
\frac{1}{2} \,
\Big[ P_{\scriptscriptstyle R1} \cos^2\theta_{e\mu}
+ (1 -P_{\scriptscriptstyle R1})\sin^2\theta_{e\mu} \Big] :
\Big[ P_{\scriptscriptstyle R1} \sin^2\theta_{e\mu}
+ (1-P_{\scriptscriptstyle R1})\cos^2\theta_{e\mu} \Big]
\label{eq:021}
\end{equation}
leading to
\begin{equation}
{\cal S} = {\cal H} = \frac{1}{2}
\left[ \frac{1}{2}
- \frac{1}{2} (1-2P_{\scriptscriptstyle R1}) \cos(2\theta_{e\mu})
\right]
\hskip0.25in , \hskip0.25in
{\cal K} = \frac{1}{6} + \frac{1}{3}
\left[\frac{1}{2}
- \frac{1}{2} (1-2P_{\scriptscriptstyle R1}) \cos(2\theta_{e\mu})
\right]
\label{eq:022}
\end{equation}
When we neglect the energy dependence of the detection rates,
there is a region which satisfies the three detection rates
within $2\sigma$ uncertainties
\begin{equation}
0.40 \alt
\left[ \frac{1}{2}
- \frac{1}{2} ( 1 -2 P_{\scriptscriptstyle R1} ) \cos(2\theta_{e\mu}) \right]
\alt 0.70.
\end{equation}
The allowed region is similar to the case in
which both neutrinos are pseudo-Dirac (see Eq.(\ref{eq:013aa}))
since in both cases there is only one adiabatic MSW transition region.

In case (II) there are two MSW resonance regions,
but in the region $R2$ the transition is extremely non-adiabatic,
i.e.
$P_{\scriptscriptstyle R2} \simeq 1$.
The neutrino fluxes at the earth depend on the
Landau-Zener
transition probability $P_{\scriptscriptstyle R1}$.
By assuming a mixing
between $\nu_1^{\mu}$ and $\nu_e$
with mixing angle $\theta_{e\mu} $, we have
\begin{equation}
\nu^e_{\scriptscriptstyle L} : \nu^{\mu}_{\scriptscriptstyle L} =
\Big[ P_{\scriptscriptstyle R1} \cos^2\theta_{e\mu}  +
( 1 - P_{\scriptscriptstyle R1} ) \sin^2\theta_{e\mu} \Big] :
\frac{1}{2}
\Big[  P_{\scriptscriptstyle R1} \sin^2\theta_{e\mu}
+ ( 1 - P_{\scriptscriptstyle R1} ) \cos^2\theta_{e\mu} \Big]
\label{eq:023}
\end{equation}
leading to
\begin{equation}
{\cal S} = {\cal H} = \frac{1}{2} - \frac{1}{2}
( 1 - 2 P_{\scriptscriptstyle R1} ) \cos(2\theta_{e\mu})
\hskip0.25in , \hskip0.25in
{\cal K} = \frac{1}{12} + \frac{11}{12}
\left[ \frac{1}{2} - \frac{1}{2}
( 1 - 2 P_{\scriptscriptstyle R1} )
\cos(2\theta_{e\mu}) \right]
\label{eq:024a}
\end{equation}
When we neglect the energy dependence of the detection rates,
we find a region which satisfies the three detection rates
within $2\sigma$ uncertainty as
\begin{equation}
0.24 \alt
\left[ \frac{1}{2}
- \frac{1}{2} ( 1 -2 P_{\scriptscriptstyle R1} ) \cos(2\theta_{e\mu})
\right]
\alt 0.35.
\end{equation}
This allowed region is similar to the case of the ordinary MSW result
(see Eq.(\ref{eq:ordmsw})).
Note that this case is different from the previous one
and the case of both neutrinos being pseudo-Dirac
because the electron neutrino flux is not depleted in half.

In summary,
the allowed regions
in the mass-mixing angle parameter space which are obtained from the
SAGE, Homestake and Kamioka experiments are very
different depending on the particle content of the neutrino sector.
For example, the region allowed when both $\nu_e$ and $\nu_{\mu}$ are
pseudo-Dirac neutrinos is very different from the region allowed by the
analysis based on the usual two generation MSW effect.
Consequently, in order to pine down the values of
$\Delta m_{e\mu}^2$ and  $\theta_{e\mu}$ from future
solar neutrino experiments,  it is necessary to have
a complete understanding of the neutrino sector,
in particular whether sterile neutrinos actually exist or not;
if they do, what their nature would be and so on.
Finally, we conclude with short comments on the apparent
atmospheric neutrino puzzle
and the neutrinoless double beta decay.
First, the atmospheric neutrino puzzle is that
\cite{ATMO}
\begin{equation}
{
\Phi_{\nu_\mu}^{\rm obs} \, / \, \Phi_{\nu_\mu}^{\rm cal}
\over
\Phi_{\nu_e}^{\rm obs} \, / \, \Phi_{\nu_e}^{\rm cal}
} =
\left\{ \begin{array}{ll} \displaystyle \vphantom{\Bigg|}
0.65 \pm 0.08 \pm 0.06
& \displaystyle
\mbox{ Kamioka }
\\ \displaystyle \vphantom{\Bigg|}
0.64 \pm 0.09 \pm 0.12
& \displaystyle
\mbox{ IMB }
\end{array} \right.
\label{eqATMO}
\end{equation}
where
$ \Phi_{\nu_{e},\nu_{\mu}}^{\rm obs} $
and
$ \Phi_{\nu_{e},\nu_{\mu}}^{\rm cal} $
are the observed and calculated
fluxes of atmospheric neutrinos, respectively.
This puzzle can easily be solved in the scenario in which both
$\nu_e$ and $\nu_{\mu}$ are pseudo-Dirac, as mentioned in Ref.\cite{KLN91},
or in the scenario (II) discussed above as long as the constraints
$\Delta m_e^2 \alt 10 ^{-7} {\rm eV}^2$ and
$ 10^{-4} {\rm eV}^2 \alt \Delta m_{\mu}^2 \alt 10^{-2} {\rm eV}^2 $
are met.
The upper limits are both due to the cosmological argument \cite{BD90}
and the lower limit for
$ \Delta m_{\mu}^2 $
is necessary in order to have an oscillation length much shorter than the
radius of the earth.
In these scenarios, $\nu_{\mu}$ is depleted in half
simply because of its pseudo-Dirac nature.
Furthermore,
a value
$ \Delta m_{\mu}^2 \sim 10^{-4} {\rm eV}^2 $,
which correspond to an oscillation length
equal to the earth diameter for
$ E \sim 500 \, {\rm MeV} $,
could explain the observed suppression of the flux
of low-energy muon neutrinos
and a value between 0.5 and 1 for the ratio given in Eq.(\ref{eqATMO}).
Note that,
in the case in which both neutrinos are pseudo-Dirac,
$\nu_e$ is not depleted because the oscillation length is much longer than the
radius of the earth due to the cosmological limit mentioned above.
Secondly,
as already discussed in the past \cite{NBBD},
neutrinoless double beta decay rates are naturally suppressed because they
become proportional to
${m^e_{\scriptscriptstyle L}}^2$,
for the pseudo-Dirac neutrinos that we have discussed.
This implies that non-observation of neutrinoless double beta decay cannot
automatically lead to the conclusion that the electron neutrino is a Dirac
particle.
\vskip2cm
\centerline{\Large\bf Acknowledgments }
\bigskip
C.G. wishes to thank the Department of Physics and
Astronomy, The Johns Hopkins University, and
C.W.K. wishes to thank the INFN, Sezione di Torino, and il Dipartimento di
Fisica Teorica, Universit\`a di Torino, for the hospitality extended to them
while part of
this work was performed.

\newpage
\noindent{\Large\bf Figure Captions}

\begin{description}

\item[Fig.1]
Effective masses squared in matter for the energy
eigenstates $\nu_1$, $\nu_2$, $\nu_3$ and $\nu_4$ as
functions of the matter density $\rho$.

\item[Fig.2]
$ \Delta m_{e\mu}^2$--$\sin^2(2\theta_{e\mu}) / \cos(2\theta_{e\mu})$
plots.
The area inside the dotted line is the allowed region in the
standard MSW effect, whereas the shaded area is the allowed region
in the pseudo-Dirac case.
The transition at $R2$ becomes non-adiabatic
in the region below the solid line in the upper right corner.

\end{description}
\end{document}